%% file: dm_substructure_paper_arXiv.tex
\documentclass[11pt,letter,floatfix]{article}
\pdfoutput=1
\usepackage{jcappub}
\usepackage{float}
\usepackage{amssymb}

\usepackage{gensymb}
\usepackage{caption}
\usepackage{graphicx}
\usepackage{amsmath}
\usepackage{wasysym}
\usepackage{verbatim}
\usepackage[utf8]{inputenc}
\usepackage{adjustbox}
\usepackage{graphicx}

\begin{document}
\title{Searching for Dark Matter Sub-structure with HAWC}

\emailAdd{lundeenj@msu.edu}
\emailAdd{jpharding@lanl.gov}
\input{authors027.tex}

\date{\today}
	\abstract{
		Numerical simulations show that the dark matter halos surrounding galaxies are expected to contain many over-densities or sub-halos.  The most massive of these sub-halos can be optically observed in the form of dwarf galaxies.  However, most lower mass sub-halos are predicted to exist as dark dwarf galaxies: sub-halos like dwarf galaxies with no luminous counterpart.  It may be possible to detect these unseen sub-halos from gamma-ray signals originating from dark matter annihilation.  
		The High Altitude Water Cherenkov Observatory (HAWC) is a very high energy (500 GeV to $>$100 TeV) gamma ray detector with a wide field-of-view and near continuous duty cycle, making HAWC ideal for unbiased sky surveys. We perform a search for gamma ray signals from dark dwarfs in the Milky Way halo with HAWC. We perform a targeted search of HAWC gamma-ray sources which have no known association with lower-energy counterparts, based on an unbiased survey of the entire sky. With no sources found to strongly prefer dark matter models, we calculate the ability of HAWC to observe dark dwarfs. We also compute the HAWC sensitivity to potential future detections for a given model of dark matter substructure.  Assuming thermal dark matter, we find the corresponding J-factor of a dark dwarf required to reach the HAWC detection criterion is $5.79\times 10^{20}~\mathrm{GeV^2~cm^{-5}~sr}$ for one particular set of dark matter assumptions.  HAWC is found to be able to competitively constrain dark matter annihilation from discovered halos with J-factors on the scale of $ 10^{19}~\mathrm{GeV^2~cm^{-5}~sr}$ or greater, with better constraints obtained on dark matter models with ~10 TeV masses and sources that transit overhead.
	}
\maketitle

\section{Introduction}
The mass of the known universe is dominated by dark matter.  Its effects can be seen in phenomena such as galactic velocity dispersion and gravitational lensing through galaxy clusters \cite{dark_evidence}.   These effects cannot be explained by normal, luminous matter within the known laws of gravity.  This leads to the hypothesis that halos of dark matter surround the luminous components, providing the mass necessary to explain these effects \cite{wimp}.  However, the composition of the dark matter remains unknown.

One popular class of dark matter candidates are weakly interacting massive particles or WIMPs.  This hypothetical particle interacts with Standard Model particles via a weak-scale force \cite{wimp}.  Under this hypothesis, dark matter annihilation can produce Standard Model particles through weak interactions. These particles, in turn, produce gamma rays mainly via pion decay , but also through inverse Compton scattering of photons off produced electrons and positrons pairs. Searching for dark matter through these photonic signals is known as an indirect dark matter search. 

 Searches for dark matter signals typically focus on regions known to be heavily dark matter dominated, such as the dwarf spheroidal galaxies surrounding the Milky Way \cite{dwarf}.  These dwarf galaxies form as a result of dark matter substructure (sub-halos) within the Milky Way halo.  These sub-halos form a gravitational nucleus around which stars can form, leading the formation of a dwarf galaxy. However, models of the evolution of the early universe predict there to be far more of these sub-halos than currently observed. 
 Simulations of dark matter show that many more sub-halos are expected to exist
than are optically observed \cite{satellites}.  However, lower-mass sub-halos are not expected to have a luminous counterpart due to lack of gravitation power and processes that suppress star formation \cite{dwarf_back}~\cite{completeness}. 
We refer to this unseen substructure as consisting of "dark dwarfs" - nearby dark matter halos with no luminous counterpart that would escape current optical observation.
 It should be noted that the inclusion of normal matter in these simulations reduces the amount of expected substructure, but non-luminous sub-halos are still expected \cite{notsolumpy}. 

Dark matter annihilation in these unseen dwarfs could produce gamma rays.  Therefore, it may be possible to detect dark matter signals from these previously unobserved sub-halos by performing an all-sky survey for gamma-ray signals. 

In this paper, we first investigate resolved HAWC gamma-ray sources that might be associated with dark matter sub-halos.  Finding no sources that significantly preferred the dark matter hypothesis, we calculate characteristic upper limits on dark matter annihilation for an unbiased sample of the HAWC sky.  Using this information in combination with simulations of dark matter substructure formation, we then compute the HAWC sensitivity to detections of dark dwarf signals.
\section{Dark Matter and Gamma Rays}

The expected differential photon flux (per unit energy) from a dark matter halo is described by the following equation:

\begin{equation}
\frac{d\Phi}{dE}=\frac{J \langle \sigma v \rangle}{8 \pi M^2}\frac{dN(M,channel)}{dE}
\label{dm}
\end{equation}
where $\langle \sigma v \rangle$ is the velocity-weighted annihilation cross section and $M$ is the dark matter mass.  The $J$-factor is defined as:
\begin{equation}
J= \int\int \rho_{dm}^2(l,\Omega) dl d\Omega \: ,
\label{j}
\end{equation}
 an integral of the squared dark matter mass density profile over the line of sight and solid angle of the observation.

\begin{figure}
 	\begin{center}
 		
 		\includegraphics[width=.6\textwidth]{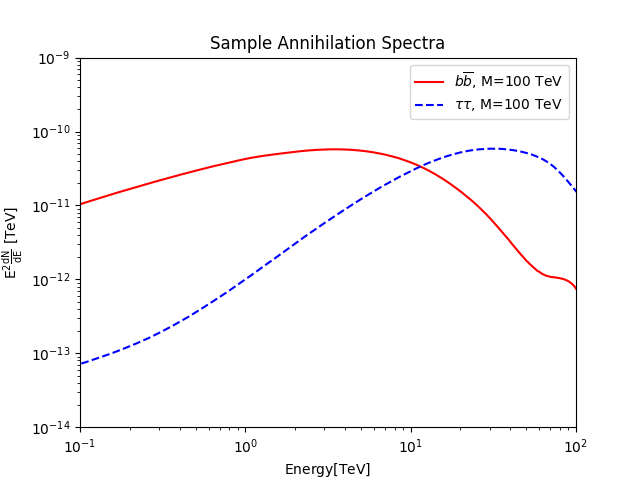}
 		\caption{Example dark matter energy spectra for two annihilation channels ($b \overline{b}$ and $\tau^+ \tau^-$).  The spectra assume a mass of 100 TeV and cut off sharply at this energy.  These are used in Eq.~\ref{dm} and they show the same characteristic shape.
 		}
 		\label{samplespec}
 		
 	\end{center}
 \end{figure}

The quantity $\frac{dN}{dE}$ is the gamma-ray spectrum from a single dark matter annihilation \cite{dark_back}. We use PYTHIA 8.2 to calculate this function by simulating dark matter annihilation and recording the number of gamma rays produced;  
our models assume 100 \% branching ratios of dark matter into individual standard model particle channels \cite{pyth}.  We include the $b \overline{b}$ channel since this has been extensively studied in other experiments as well the $\tau^+ \tau^-$ channel because this is the heaviest solely leptonic channel available.  An example of dark matter spectral shapes, showing the characteristic hard energy cut-off at the dark matter mass, is shown in Fig.~ \ref{samplespec}.  

\section{HAWC}

The High Altitude Water Cherenkov (HAWC) detector is a gamma-ray observatory located at Sierra Negra, Mexico. Consisting of an array of 300 water Cherenkov detectors and covering an area of 22,000 $\mathrm{m^{2}}$, it is able to detect gamma rays by the air showers they produce in the atmosphere. Air showers are created when charged particles or gamma rays interact with Earth's atmosphere, producing cascades of lower-energy particles.  HAWC detects these secondary particles by the Cherenkov light they produce while passing through the water Cherenkov detectors, which is then detected with four photomultiplier tubes (PMTs) mounted at the bottom.  Timing information is then used to reconstruct angle of arrival, and the distribution of charges is used  \cite{hawcback}.  

The majority of air showers detected by HAWC come from charged cosmic rays (mostly protons).  Therefore, cuts are applied to separate the hadronic events from the gamma rays before the data is analyzed. With cuts, 99\% of hadronic events are rejected at the highest energies \cite{hawcback}.  To estimate the remaining background after cuts, we use a technique called direct integration. Events are integrated within two hours in right ascension of a point to estimate the associated background \cite{hawcback}.  Our signal at each point in the sky is then the number of excess gamma rays above this estimated background level. Due to atmospheric effects HAWC is sensitive to the declination at which a source transits, with the best sensitivity obtained from points that transit overhead \cite{hawcback}.

HAWC is sensitive to gamma rays between 500 GeV and $>$100 TeV and is well-suited for detecting signals from multi-TeV dark matter masses.  In addition, HAWC operates on a near-continuous duty cycle with a wide field-of-view that makes it ideal for performing survey-style observations \cite{hawcback}. HAWC does not need to be oriented to observe a source and collects data from 2/3 of the sky each day.  With these properties, we perform an unbiased search for dark dwarfs.  

\begin{figure}
\begin{minipage}{.5\textwidth}
		\begin{tabular}{c|c}
			Bin  &  $f_{hit}$ bounds (\% of PMTs hit) \\
			\hline
			1 & 6.7-10.5 \\
			2 & 10.5-16.2\\
			3 & 16.2-24.7\\
			4 & 24.7-35.6\\
			5 & 35.6-48.5\\
			6 & 48.5-61.8\\
			7 & 61.8-74.0\\
			8 & 74.0-84.0\\
			9 & 84.0-100.0\\
		\end{tabular}

\end{minipage}
\begin{minipage}{.5\textwidth}
 \includegraphics[width=\textwidth]{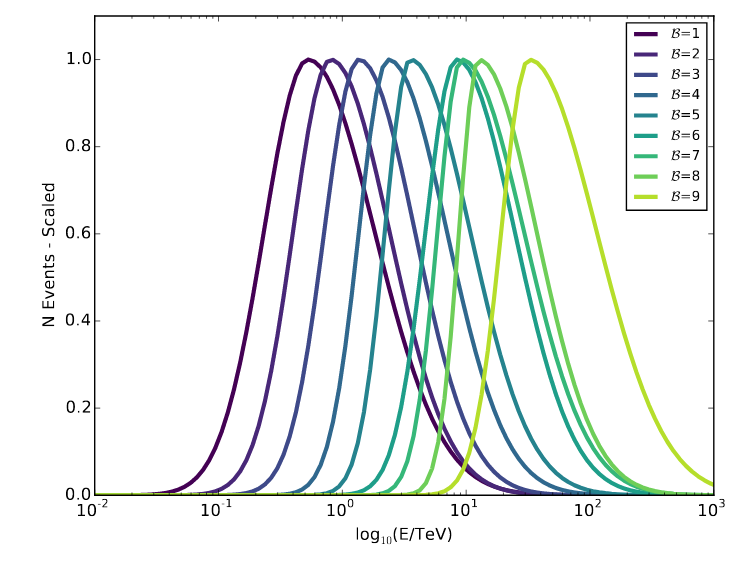}
 \end{minipage}
\caption{Bin definitions for the $f_{hit}$ binning scheme used for HAWC data in this paper (left) and scaled histogram of the reconstructed energy distribution of each bin (right). The right figure is taken from Fig.~3 of Ref.~\cite{hawcback} and was generated assuming a power law spectrum $\frac{d \Phi}{dE} \sim E^{-2.66}$ for a single transit of a source at a declination of 20 degrees.  All distributions have been normalized to have unity amplitude for visualization purposes.  }
\label{fig:bins}
\end{figure}

\begin{figure}
	\centering

	\includegraphics[width=.45\textwidth]{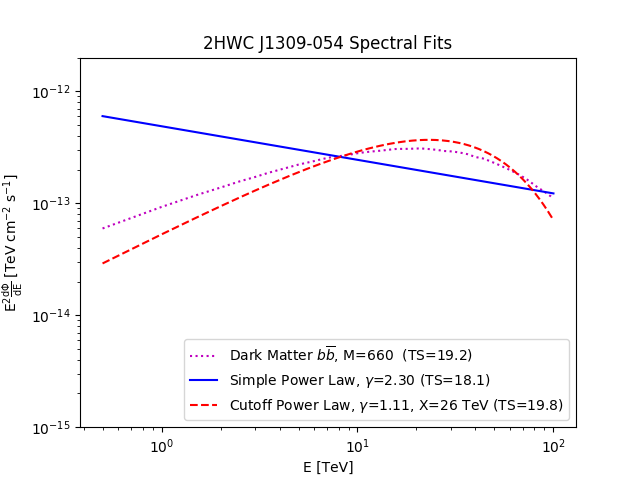} 
	\includegraphics[width=.45\textwidth]{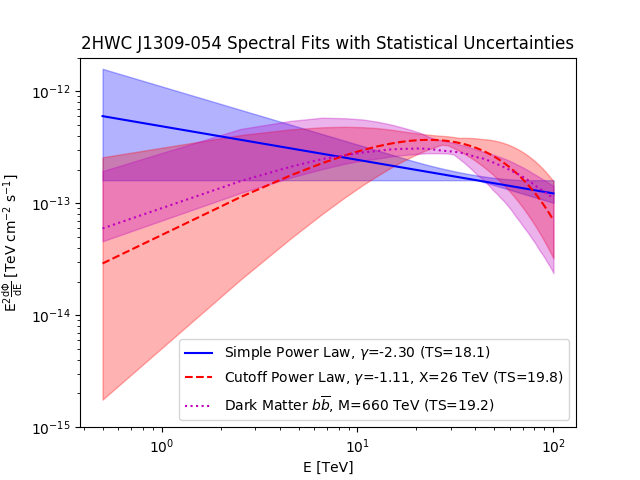}\\ 
	\includegraphics[width=.45\textwidth]{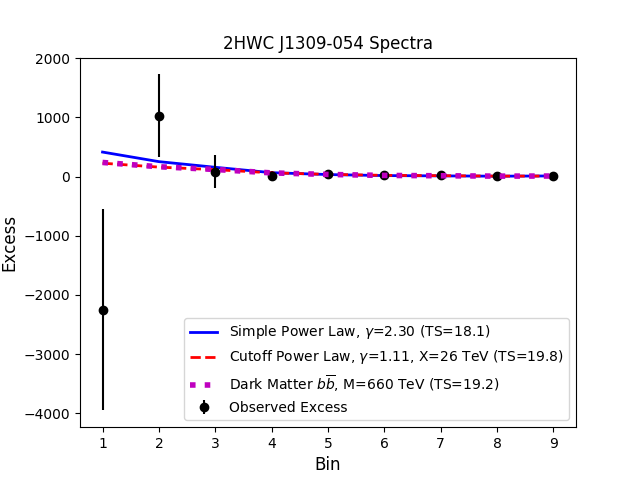}
	\includegraphics[width=.45\textwidth]{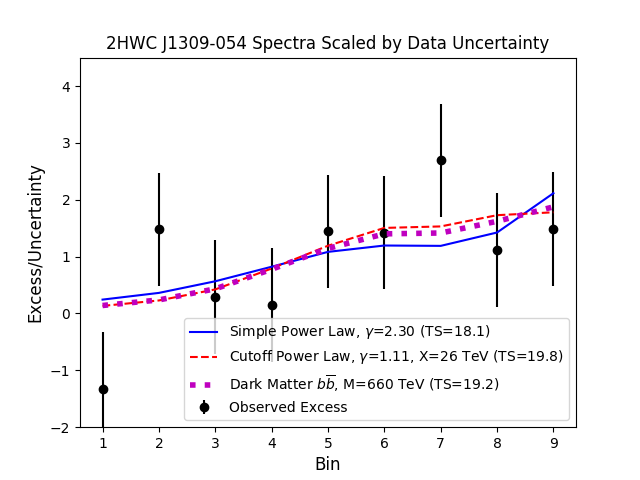}
	\caption{Sample best-fit spectra plotted both in energy (top) and $f_{hit}$ bins (bottom) to one of the unassociated sources (2HWC J1309-054) considered in this analysis for its best-fit simple power law, cut-off power law, and dark matter spectrum.  While the spectra appear distinct in energy-space, they are of comparable goodness of fit to the data in $f_{hit}$ bins.  The top right plot also shows the statistical uncertainties of the best fits as correspondingly-colored shaded regions, indicating that the range of allowed fits overlaps substantially.  We show the $f_{hit}$ bin fits both as terms raw counts (left) and scaled by the estimated error in the data (right).  The $f_{hit}$ bin plots (bottom) are calculated summing events over a fixed angular disk in the sky rather than the full spatial-likelihood calculation. These are for visualization purposes only. The shown best-fit spectra and TS (Eq.~\ref{ts})values are those from the full spatial-likelihood calculation. }
	\label{nsamp}

\end{figure}

The algorithms used in this analysis have relatively broad energy resolution (see Fig.~\ref{fig:bins}).  Therefore, rather than directly reconstruct particle energy, HAWC uses a forward folding method to fit spectra \cite{hawcback}.  Events are binned not in energy but in $f_{hit}$, the fraction of available PMTs that detect signal in a shower event (see Fig.~\ref{fig:bins} for bin definitions and their corresponding energy distributions).
The observable spectrum in $f_{hit}$ space is sensitive to the energy distribution.   To fit energy spectra, we convolve the HAWC $f_{hit}$ response to gamma-ray spectra with different spectral parameters to convert the spectra to $f_{hit}$ space.   Using these $f_{hit}$ spectra, we find the energy spectrum which best matches the measured $f_{hit}$ spectrum with maximum likelihood methods \cite{hawcback}.  Some examples of fits to such a forward folded spectral fitting are shown in Fig.~\ref{nsamp} for a resolved HAWC source; see Sec.~\ref{search} for details on the spectra used.  Note that future HAWC analyses will use recently developed algorithms to reconstruct energy directly.

HAWC fits the energy spectra of its data using a maximum likelihood ratio test with respect to background.  For an assumed model, we fit to maximize the test statistic (TS) defined in Eq.~\ref{ts}. 
 \begin{equation}
 \mathrm{TS}=-2 \log{\big(L_{0}/L \big)}
 \label{ts}
 \end{equation}
 where $L_{0}$ is the likelihood for the null hypothesis and $L$ is the likelihood for signal.  See Appendix A of Ref.~\cite{dwarf} for details on how the likelihoods are evaluated for HAWC data.

\section{Searching for Dark Dwarfs in HAWC Data}
\subsection{Search Method}
\label{search}
We begin our analysis by searching 760 days of HAWC data for gamma-ray excesses that could potentially originate from dark dwarfs.  We consider all sources with a TS (Eq.~\ref{ts}) of 25 or greater in the 2HWC catalog, had a galactic latitude, $\lvert b \rvert > 5 \degree$ and have no known astrophysical counterpart reported \cite{cat}.  Because these sources had no known normal-matter association and are so far off the Galactic plane, each one is a potential dark dwarf.  Note that although all of these sources had a TS greater than 25 in the 2HWC catalog, three of these sources have TS values less than 25 in our new dataset of 760 days.  No additional sources  with TS greater than 25 were found in the 760 day dataset.  Therefore, we only consider these four sources in this analysis. 

The sources are listed in Table \ref{table:sources}, where RA is the right ascension, Dec is the declination, and the radius is the angular size of the disk source hypothesis used in the 2HWC catalog. A radius of zero indicates a point source (smaller than the HAWC angular resolution).  Sources with larger angular extent would indicate dark dwarfs that are either close to Earth or have large spatial extent.  Because we use a different data set than the 2HWC catalog (which consisted of 507 days of data rather than 760), the maximum TS values for the power law hypothesis will differ from those reported in the catalog consistent with fluctuations from additional data \cite{hawcback}.  If the charged cosmic ray background fluctuates to a higher value, but not the signal, this can lower the TS relative to that reported in the catalog, which is the case for three of our four sources (see Table~\ref{table:results} in the following section.)

\begin{table}
\begin{center}
\begin{tabular}{c|c|c|c}
Name  &  RA ($\degree$)   &   Dec ($\degree$)   &   radius ($\degree$)  \\
\hline
2HWC J0819+157  &  124.98 & 15.79  & 0.5\\
2HWC J1040+308  &  160.22 & 30.87  & 0.5 \\
2HWC J1309-054  &  197.31 & -5.49 & 0\\
2HWC J1829+070  &  277.34 & 7.03  &  0 \\
\end{tabular}
\end{center}
\caption{Location and spatial data for potential dark dwarfs considered in this analysis. Radius refers to the angular size of the disk source hypothesis (zero corresponding to a point source).  All four of these sources were seen in the 2HWC catalog with maximum of TS~$>$~25 \cite{cat} and no additional resolved sources were found in the dataset used in this analysis.  Only one of these sources (2HWCJ 1040+308) still remains at TS~$>$~25 in the 760 day dataset.}
\label{table:sources}
\end{table}

 To test these sources for possible dark matter signals, we compare fits of the spectra predicted by Eq.~\ref{dm}, to those predicted by astrophysical spectra (Eq.~\ref{pl} and Eq.~\ref{cu}).  
\begin{equation}
\frac{d\Phi}{dE}=\Phi_{0}(E/E_{0})^{-\gamma} 
\label{pl}
\end{equation}
\begin{equation}
\frac{d\Phi}{dE}=\Phi_{0}(E/E_{0})^{-\gamma} e^{-E/X}
\label{cu}
\end{equation}
Eq.~\ref{pl} is a simple power law where flux is proportional to energy raised to a power $\gamma$, while Eq.~\ref{cu} is a cut-off power law with the addition of an exponentially decaying factor, with the cut-off energy $X$.  
In both cases $E_0$ is the pivot energy, which we fix to 7 TeV .  This choice of parameterization has no effect on the actual spectral shape but does effect the correlation between the analysis bins, and we choose this value in order to minimize the correlation (see Ref. \cite{hawcback}).
In the case of the dark matter hypothesis, we do not assume density profiles and therefore do not independently compute $J$ (Eq.~\ref{j}). Instead, we treat $\langle \sigma v \rangle J$ as a single free parameter, with dark matter mass being the other.

For each of our models, the null hypothesis is that the spectrum is background-only. Each alternative hypothesis is tested separately and the TS measured.  We check to see if the dark matter hypothesis produces a TS at least has high as the most favored (highest TS) power law.   If this is the case, further analysis with a closer examination of the likelihood distributions will be used to determine if the dark matter hypothesis is statistically favored, and the observed emission a possible detection of a dark dwarf.
 
\subsection{Fit results}

\begin{figure}[H]
	\centering
	\includegraphics[width=.45\textwidth]{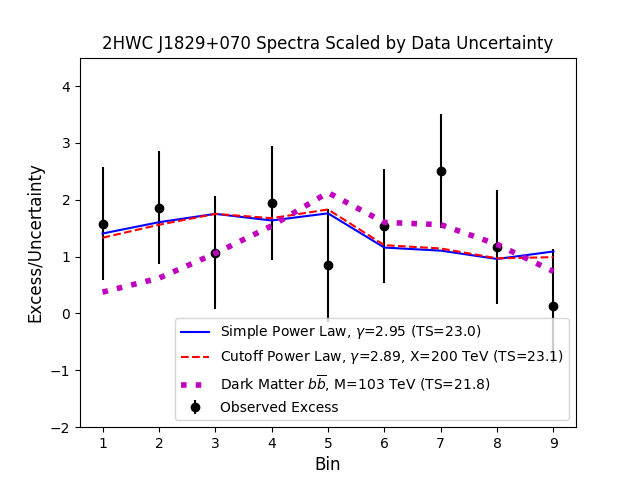}
	\includegraphics[width=.45\textwidth]{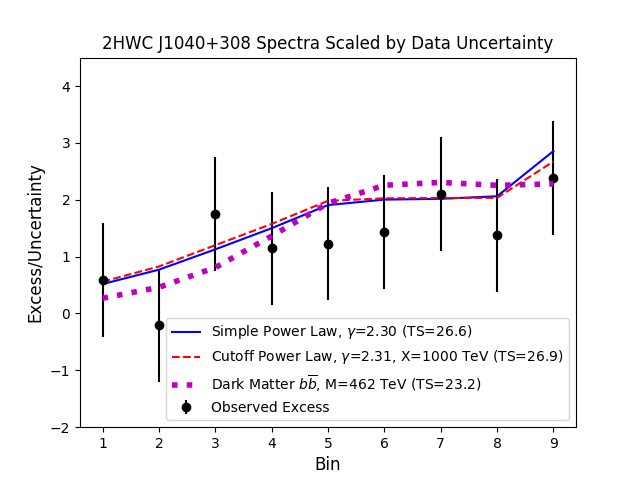}
	\\
	\includegraphics[width=.45\textwidth]{2HWCJ1309054_binned_data_errors.png}
	\includegraphics[width=.45\textwidth]{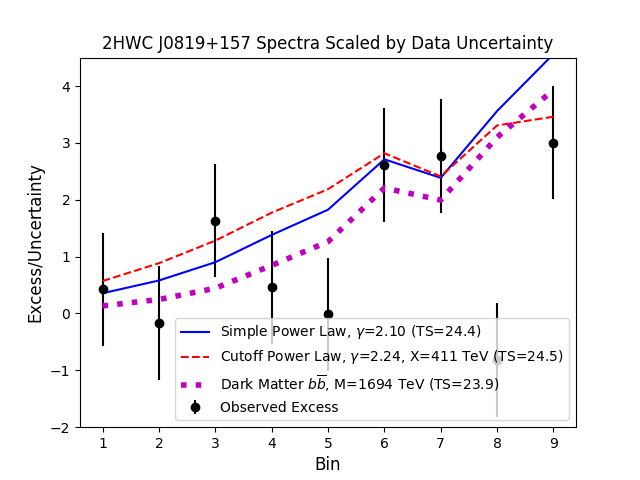}
	\caption{Best-fit results in $f_{hit}$ bins for the four targeted sources showing the three fitted spectra and TS values.  The counts in these plots are calculated and normalized as in the lower right plot of figure 4. The counts do not reflect the full spatial-likelihood analysis of HAWC and are for visualization purposes. The best-fit TS numbers are from the full HAWC likelihood calculation. Each spectrum is scaled by the estimated uncertainty in data (as was the case in the lower-left plot of Fig. \ref{nsamp}), showing that each best-fit hypothesis is comparable in goodness of fit. We assume the same spatial hypotheses for each as the 2HWC catalog (see Table.~\ref{table:sources}) \cite{cat}}
	\label{fitresults}
\end{figure}

For each source, we fit a power law, cut-off power law and dark matter spectrum. The fits are performed using the Multi-Mission Maximum Likelihood framework (3ML) \cite{3ml} \cite{3mlgit}. The dark matter fits have two free parameters: the $J \langle \sigma v \rangle $ pre-factor and $M$, the dark matter mass.  We vary spectra generated assuming different $M$ and annihilation channels calculate the $J \langle \sigma v\rangle $ that maximize the likelihoods in each case.  The best fits for the spectra are shown in Fig.~\ref{fitresults} and summarized in Table \ref{table:results}.; only the dark matter spectrum with the highest TS is plotted in each case.

 We observe no significant improvement in the TS for the dark matter spectra compared to the power laws. In each case, the fits are either comparable or slightly worse (lower TS) for the dark matter hypothesis. It should also be noted that the TS values are nearly indistinguishable between the two power law hypotheses, indicating that the sources lack sufficient statistical power to differentiate even nested spectral shapes. 
 Therefore, we cannot declare the dark matter hypothesis to be favored. 
 However, it is important to note that these results do not exclude the possibility of dark matter signals from the surveyed sources; they simply do not favor the dark matter hypothesis over other astrophysical spectra. 

\begin{table}
\begin{center}
\begin{tabular}{c|c|c|c|c|c}
	Name  &  TS  (PL)    & TS (CU)  & TS (DM) & $\Delta$ TS (DM-PL)  & $\Delta$ TS (DM-CU)  \\
	\hline
	2HWC J0819+157  &  24.4 & 24.5 & 23.9 & -0.5 & 0.6 \\ 
	2HWC J1040+308  &  26.6 & 26.9  &  23.2 & -3.3 & -3.7\\
	2HWC J1309-054  &  18.1 & 19.8  & 19.2 &  1.1 & -0.6\\
	2HWC J1829+070  &  23.0 & 23.1    & 21.8 & -1.2 & -1.3\\
\end{tabular}
\end{center}
\caption{Summary of test statistics (TS) and $\Delta$ TS for the three flux models considered in this paper: power law (PL), cut-off power law (CU), and the best-fit dark matter spectrum (DM). See Fig.~\ref{fitresults} for details on best-fit spectra for each source.}
\label{table:results}
\end{table}

\section{Characteristic Upper Limits Across The Sky}
\label{all-sky}
With no clear detections of dark matter sub-halos in the HAWC data, we proceed to estimate characteristic upper limits on dark matter annihilation within the HAWC field-of-view.   Characteristic upper limits refer to the average dark matter $\langle \sigma v \rangle$ which HAWC could exclude at 95\% confidence as a function of declination, assuming a dwarf with known J-factor were discovered at that declination and no gamma-ray excess were observed. 
We estimate our characteristic limits by finding the average flux which could be excluded by HAWC at 95\% confidence at each declination. If a dark dwarf were discovered by another instrument, these characteristic limits could be used to give the expected value of the corresponding dark matter annihilation upper limit.  Note that for any one given discovery of a dwarf, fluctuations in the background at that location would vary the corresponding limit from the characteristic limit.

To calculate the characteristic upper limits we use grid for $0\degree$ $<$ RA $<$ $360 \degree$ in steps of one degree and $-10\degree$ $<$ Dec $<$ $55\degree$ in steps of five degrees.  We chose the declination spacing of five degrees to account for the HAWC detector response to different declinations, which does not substantially change on the scale of five degrees.  We chose the right-ascension due to the HAWC point spread function (PSF), which has a width of approximately one degree in the lowest $f_{hit}$ bin \cite{hawcback}.  A spacing of one degree ensures each sampled point is roughly independent (little overlap in the PSF) while still obtaining a representative sample of points at each declination.  We perform these calculations assuming point source hypotheses since the expected characteristic size of a dwarf is on the scale of the HAWC PSF width, although some dark dwarfs may have slightly larger extent \cite{ext1} \cite{ext2}. Note that the points analyzed in Sec. \ref{search} were drawn from a search of all points in the entire HAWC sky \cite{cat}, and all possible sources of dark matter have already been identified so there is no chance an additional TS~=~25 excess will be missed in this analysis.  

\begin{figure}[H]
	\begin{center}
		
		\includegraphics[width=.7\textwidth]{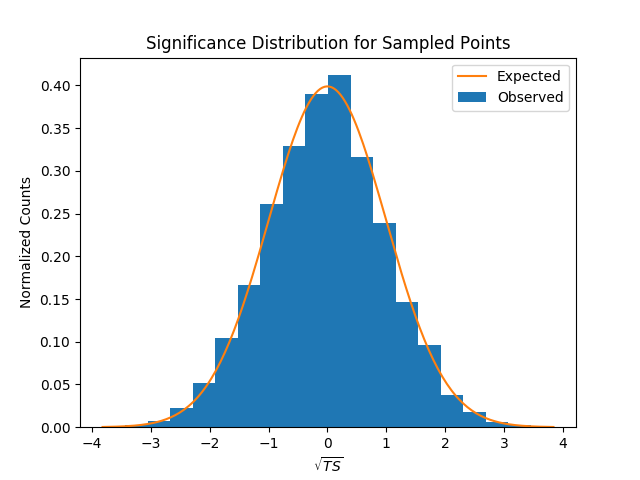}
		\caption{Significance (as defined in the 2HWC catalog \cite{cat}) distribution of the grid points from all strips of declination used in the characteristic limit calculations (Observed) superimposed over the distribution expected from background-only (Expected).  The likelihoods used for this plot were calculated assuming a 10 TeV, $ b \overline{b}$ channel dark matter spectrum.  The shape is roughly consistent with background, indicating no excess remains after resolved sources from the catalog are removed, and that the sample is an unbiased representation of the remaining sky.}
		\label{samplehist}
	\end{center}
\end{figure}

To avoid contamination from luminous matter gamma-ray emission, we exclude any points within five degrees of known luminous matter gamma-ray sources reported in the 2HWC catalog \cite{cat}, as well as the Galactic plane.  As the HAWC PSF has a width of approximately one degree at the lowest energies (and narrows at higher energies), we choose five degrees in order to exclude both the main emission and the tails due to the PSF as well as emission from highly extended sources \cite{hawcback}. An example significance distribution is shown in Fig.~\ref{samplehist} and is consistent with background, indicating we have removed significant source contamination and are using a sufficiently unbiased sample of the sky. Then, we fit each of our sampled grid points with spectra from a range of assumed dark matter masses and annihilation channels, assuming point sources. As we expect the dark dwarfs to be relatively low-mass and/or far away, we fit each pixel with a point source model.
Since we do not assume any given dark matter density profiles or distances to the satellites, our characteristic limits are on $J \langle \sigma v \rangle $ rather than just $\langle \sigma v \rangle$.

\subsection{Characteristic Limits}
\label{charlims}
\begin{figure}[H]
	\begin{center}
		\includegraphics[width=.9\textwidth]{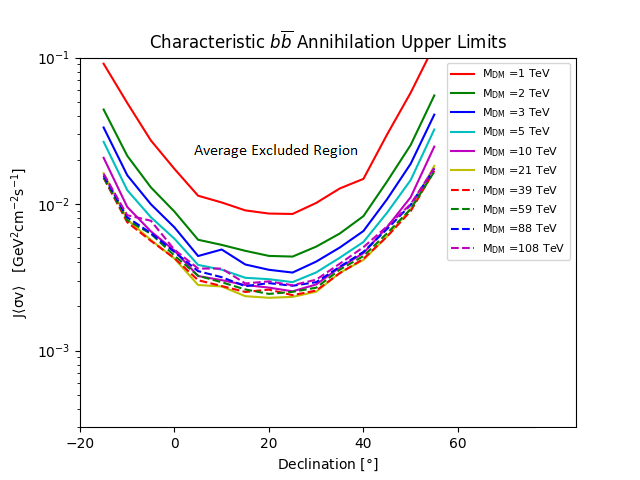}\\
		\includegraphics[width=.9\textwidth]{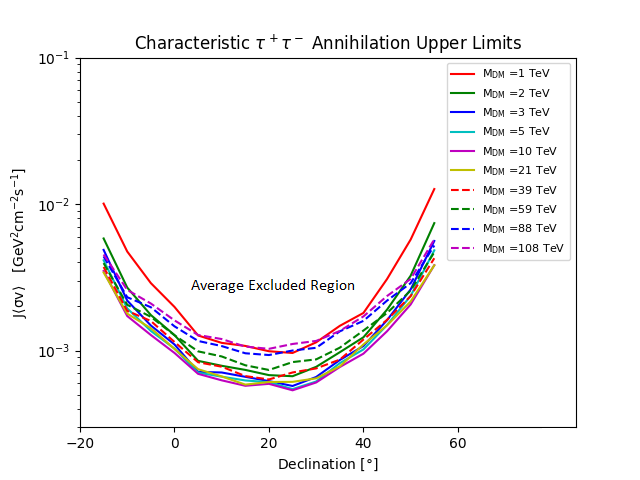}
		\caption{Characteristic upper limits on $J \langle \sigma v \rangle$ for the $b \overline{b}$ and $\tau^+ \tau^-$ channels at each declination.  The points plotted show the expected 95\% confidence level upper limit for points at each declination.  The most constraining limits are found at declinations directly overhead for HAWC.  Individual dwarf upper limits will vary depending on statistical fluctuations at their locations.}
		\label{limits}
	\end{center}
\end{figure}

Since the HAWC sensitivity is highly declination-dependent, we will report our characteristic upper limits as both a function of dark matter mass and declination, rather than showing the individual limits for each sampled point. We estimate our characteristic 95\% confidence level upper limits using the distribution of fitted fluxes. Each best-fit $J \langle \sigma v \rangle$ value in a given declination is placed in a histogram, and we select our characteristic upper limit on $J \langle \sigma v \rangle $ as that which is greater than 95\% of the best-fit values.  In practice, this is equivalent to averaging the individual limits obtained above. Our characteristic limits are plotted in Fig.~\ref{limits}.

\begin{figure}[H]
	\begin{center}
		\includegraphics[width=.9\textwidth]{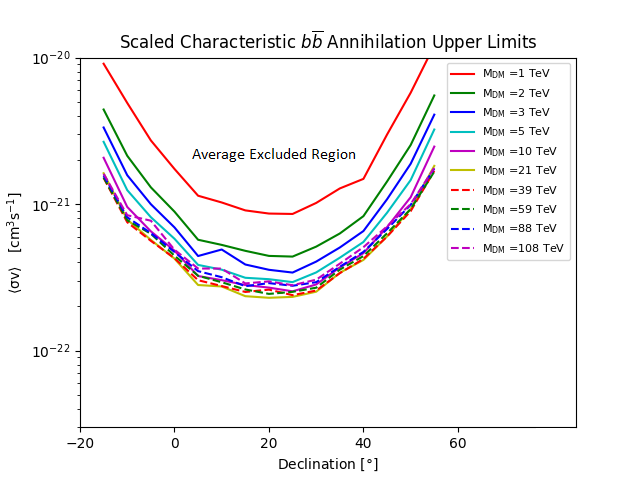}
		\includegraphics[width=.9\textwidth]{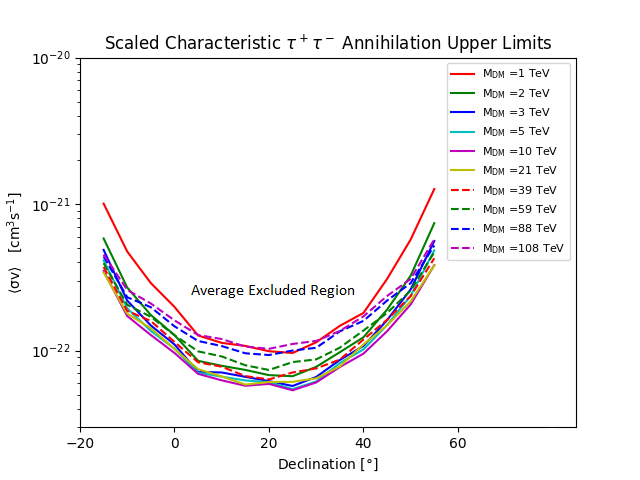}
		\caption{Characteristic upper limits on $\langle \sigma v \rangle$ for the $b \overline{b}$ and $\tau^+ \tau^-$ channels assuming a dwarf galaxy with $J = 10^{19}~\mathrm{~GeV^{2}~cm^{-5}~sr}$ were discovered at each declination.  These are obtained from scaling the limits in Fig.~\ref{limits} by this J-factor. For reference, the canonical thermal $\langle \sigma v\rangle$ value is $2.2 \times \mathrm{10^{-26}~cm^{-3}~s^{-1}}$ \cite{thermal_relic}.  Individual upper limits will depend on statistical fluctuations at the location of a discovered dwarf and its measured J-factor.}
		\label{scaled}
	\end{center}
\end{figure}

With these characteristic limits, should a new dwarf galaxy be discovered and its J-factor measured, one could find the expected $\langle \sigma v \rangle $ upper limit by scaling the limits in Fig.~\ref{limits} by the corresponding J-factor.   An example is shown in Fig.~\ref{scaled} assuming a fairly massive dwarf galaxy (with $J=10^{19}~\mathrm{~GeV^{2}~cm^{-5}}$) were discovered at each sampled declination. 
Assuming an even more massive or close dwarf with $J=8.7\times 10^{19}~ \mathrm{GeV^{2}~cm^{-5}}$ were discovered at our best declination ($20\degree$) our expected upper limits would become competitive with the HAWC combined dwarf-spheroidal upper limits.  In this case, for assuming $b \overline{b}$, M~=~10 TeV spectrum, our upper limit on $\langle \sigma v\rangle$ is $ 3\times 10^{-23}~ \mathrm{cm^{3}~s^{-1}}$ comparable to the corresponding value in the dwarf-spheroidal analysis \cite{dwarf}.  In order to become competitive with the corresponding Magic upper limits from observations of Segue 1 ($4.33\times ~10^{-24}\mathrm{cm^{3}~s^{-1}}$) \cite{magic_segue}, the J-factor would need to be at least $6 \times 10^{19} ~\mathrm{GeV^{2}~cm^{-5}}$. 

Note that the actual upper limit derived from a discovered dwarf will depend on fluctuations in the HAWC data at the dwarf location (see the analysis in Ref. \cite{dwarf}).  The limits shown here characterize the sensitivity of HAWC to dark matter emission at different declinations and show how constraining we can expect these future analyses to be.


\begin{figure}[H]
	\begin{center}
		
		\includegraphics[width=.75\textwidth]{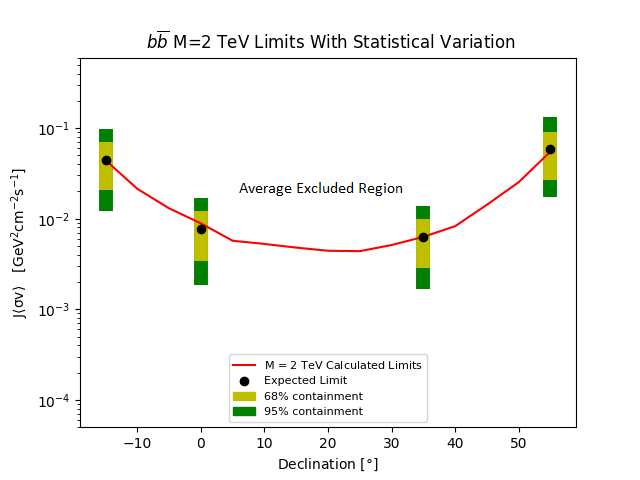}
		\caption{Statistical variation (68\% and 95\%)  on characteristic upper limits on $J \langle \sigma v \rangle$ as well as the median (expected)value, plotted as a function of declination.  These values were calculated assuming a $ b \overline{b}$ annihilation channel spectrum with a dark matter mass of 2 TeV.  The error bar size is roughly independent of declinations and spectral assumptions, so we show only this sample.}
		\label{error}
	\end{center}
\end{figure} 

\subsection{Statistical Variations and Expected Limits}
\label{stat_var}

To verify our characteristic limits, we also calculate the expected statistical variation in the upper limits.  To do so, we re-calculate the upper limits at each declination using a series of pseudo-maps generated using our measured background.   Each pseudo-map is created by injecting a simulated signal drawn from a Poisson distribution about the measured background in each bin.  We then histogram the pseudo-map limits and find the boundaries that contain 68\% and 95\% of the distribution, which demonstrates the possible statistical variation of true limits compared to the characteristic limits.  We use simulated pseudo-maps rather than our actual limit distribution because there are enough independent points in each declination bin to rigorously find the statistical variation.  The characteristic limits we calculated in Sec.~\ref{charlims} are consistent with the expected (median) pseudo-map limits, as would be expected from a background-dominated distribution. 

The resulting statistical variations and median (expected) limits are shown in Fig.~\ref{error}, superimposed over the corresponding calculated limits from Fig.~\ref{limits}.  The size of our statistical error bars was found to be roughly independent of the spectral assumptions and declinations.  Therefore, we show only a sample of error bars for select declinations and assuming a 2 TeV mass, $b \overline{b}$ dark matter spectrum.  

Our characteristic limits as a function of declination reflect the HAWC sensitivity curve \cite{hawcback}, with the most constraining limits found at points that transit directly overhead. This analysis gives an estimate of the HAWC sensitivity specifically to dark matter spectra.  In the following section, we use this estimate in conjunction with simulations of substructure to compute the HAWC sensitivity to detection of flux from an ensemble of dark dwarfs.  It should be emphasized that the results in this section show characteristic limits consistent with expected limits within a declination band.  The individual upper limits on flux from newly discovered dwarf galaxies will have statistical fluctuations as shown in Fig.~\ref{error}.


\section{HAWC Sensitivity to Modeled Dark Dwarfs}

The HAWC sensitivity estimates of the Sec. \ref{all-sky} can be applied to ensembles of dark dwarfs, not just individual dark dwarfs.  Here we calculate the HAWC ability to detect dark dwarfs based on models of dark matter sub-structure across the sky.  This will give an estimate of how likely HAWC is to observe a dark dwarf in its field-of-view.

\subsection{Modeling Dark Matter}
We use the \textsc{clumpy} (version 2015.06) package to obtain dark matter substructure models.  \textsc{clumpy} models both the main (smooth) dark matter halo as well as dark matter sub-halos (clumps) and computes the corresponding J-factor at each modeled point \cite{clumpy}.   As the substructure distribution is not perfectly constrained
by current observations, we use a set of characteristic halo models  sample of 100 Monte Carlo trials in each case. For each trial, \textsc{clumpy} creates a \textsc{healp}ix \cite{healpix} map of the J-factor associated with each point given a user defined integration angle, which we use to determine the expected halo locations.  

The exact behavior of dark matter density profiles towards the center of a halo is not well-constrained, so we use two different parameterizations consistent with numerical simulations and observational data.  Many fits to numerically simulated halos favor the Einasto density profile shown Eq.~\ref{einasto}, which is characterized by a cuspy shape towards the halo center \cite{einasto1} \cite{einasto2}.  
\begin{equation}
\rho(r)=\rho_{s} e^{\frac{-2}{\alpha}[(r/r_{s})^{\alpha}-1]}
\label{einasto}
\end{equation}
Here, $\rho_{s}$ is a normalization constant on the dark matter mass density determined by the total halo mass.  $r_{s}$ is the characteristic scale radius of the halo and $\alpha$ determines the profile's curvature. Note that in the case of the Galactic halo \textsc{clumpy} determines this constant internally using the distance from the sun to the galactic center ($R_{\odot}$) and the local dark matter density ($\rho_{\odot}$) \cite{clumpy}.  All parameter values are reported in Table \ref{table:profileparams}. 

Emperical measurements of dark matter density profiles through observation of the baryonic component of the halos may favor a Burkert profile \cite{burkert}, shown in Eq.~\ref{burkert_profile} and characterized by a cored center.  
\begin{equation}
\rho(r)=\frac{\rho_{s}}{(1+r/r_{s})(1+(r/r_{s})^{2})}
\label{burkert_profile}
\end{equation}
Here, $\rho_{s}$ and $r_{s}$ are again the density normalization and scale radius and are given the same values as in the Einasto profile for the main halo component (see Table \ref{table:profileparams}).

\begin{table}
	\begin{center}
		\begin{tabular}{c|c|c|c}
			
			$R_{\odot}$ (kpc) & $\rho_{\odot}$ ($\mathrm{GeV/cm^{3}}$)  &  $r_s$ (kpc)   &  $\alpha$    \\
			\hline
			8 &  0.4 & 15.7 & 0.17  \\  
			
		\end{tabular}
	\end{center}
	\caption{Parameters used in the assumed dark matter density profiles (Eq. \ref{einasto} and Eq. \ref{burkert_profile}). $R_{\odot}$ and $\rho_{\odot}$ are respectively the distance from the sun to the Galactic center and the local dark matter density of the solar system and determine the scale density $\rho_s$. The scale radius $r_s$ is the radius such that $\rho(r_s)=\rho_s$ and $\alpha$ determines the slope of the Einasto profile (the Burkert profile does not set the slope as a free parameter). }
	\label{table:profileparams}
\end{table}

For the simulated sub-halos, $\rho_{s}$ and $r_{s}$ are determined as a function of the total subhalo mass $M_{sub}$ using the concentration parameter formalism defined in Ref.~\cite{mcfunction}.  In this framework, the halo shape parameters are related to the total halo mass by the concentration parameter $c_{vir}$ defined as: 
\begin{equation}
	c_{vir}(M_{sub})=R_{vir}(M_{sub})/r_{-2}
	\label{cvir}
\end{equation}  
where  $r_{-2}$ is the radius at which the logarithmic slope of the halo is equal to -2 and  $R_{vir}(M_{sub})$ is the virial radius of the halo.  The virial radius is given as the following function of $M_{sub}$:
\begin{equation}
	R_{vir} (M_{sub}) = \Bigg( \frac{M_{sub}}{(4\pi/3) \Delta_{vir} \Omega_{m} \rho_c} \Bigg)^{1/3}
	\label{rvir}
\end{equation}
where the values of the constants are given in Ref.~\cite{mcfunction}.
Various models exist for the functional form of $c_{vir}(M_{sub})$ and we use the one provided by Ref. \cite{mcfunction}:
\begin{equation}
	\ln(c_{vir}) = \sum C_i \ln\Big(\frac{M_{sub}}{M_{\odot}}\Big)+\ln\Big(\frac{1}{1+z}\Big)
	\label{cm}
\end{equation}
where $M_{\odot}$ is the solar mass, $z$ is the halo redshift, and the coefficients $C_i$ are taken from Eq.~3 of Ref.~\cite{mccoefficients}.
 It should be noted that to Eq.~\ref{cm} was fitted in a simulation assuming the dark matter density follows the NFW profile \cite{mccoefficients}, and the coefficients could differ had the simulation assumed an Einasto or Burkert profile.  Recent work indicates subhalos may be more concentration than the main component \cite{moline}. The model we report here is therefore relatively conservative.
  
For a given halo of mass $M_{sub}$,  $R_{vir}$ and $c_{vir}$ are calculated and used to determine $r_{-2}$ using Eq.~\ref{cvir}.   The corresponding $r_s$ is then obtained from $r_{-2}$ depending on the functional form of the density profile. For the Einasto profile, $r_s = r_{-2}$; for the Burkert profile, $r_s \approx r_{-2}/1.52$.  
Once $r_s$ is determined for a given halo mass, $\rho_s$ is determined by normalizing the integrated density profile to the total mass.  Put in equation form:
\begin{equation}
	\rho_s = \frac{M_{sub}}{4 \pi \int_{0}^{R_{vir}} f(r)  r^{2} dr}
	\label{rhos}
\end{equation} 
where $f(r)$ is the functional form of the density profile (Eq. \ref{einasto} or Eq. \ref{burkert_profile}). 

Following the Aquarius simulation, we assume the number distribution of sub-halo masses follows a power law form ($\frac{dN}{dM_{sub}} \sim M_{sub}^{-n}$), where $n$ is the power law index, and set the minimum and maximum subhalo masses considered accordingly \cite{aquarius}.  The parameters used here are summarized in Table \ref{table:simparams} and the resulting total fraction of dark matter mass contained in substructure is then 18\%. It should be noted that the values determined here from Aquarius were obtained using a set of cosmological parameters that differ from those most recently obtained by the Plank experiment \cite{planck}.

The spatial probability distribution of the dwarfs is modeled by the Einasto profile, where $\rho_{s}$ is chosen such that the profile is normalized to unity \cite{simparams}.  See Table \ref{table:simparams} for the values used. 
Note that in all cases, we only consider dwarfs with simulated J-factors $>10^{16}~\mathrm{GeV^2~cm^{-5}~sr}$.  In both cases, the sub-halo density profiles have the same functional form as the main halo profile \cite{clumpy}.

\begin{table}
	\begin{center}
		\begin{tabular}{c|c|c|c|c}
			
			$M_{min}$ ($M_\odot$) & $M_{max}$ ($M_\odot$)  & n &  $r_s$ (kpc) & $\alpha$     \\
			\hline
			 $10^{-6}$ & $10^{8}$ & 1.9  & 199 & 0.69  \\  
		
		\end{tabular}
	\end{center}
	\caption{Parameters used in \textsc{clumpy} simulation of the Galactic substructure mass function and spatial distribution.  $M_{min}$ and $M_{max}$ are respectively the maximum and minimum subhalo masses considered (in units of the solar mass $M_\odot$) and $n$ is the slope of the power law assumed for the mass function.  The spatial distribution is modeled by Eq. \ref{einasto} with the $r_s$ and $\alpha$ values reported here and $\rho_s$ choses to normalize to unity. The values reported here are taken from Ref.~\cite{aquarius} and Ref.~\cite{simparams}.}
	\label{table:simparams}
\end{table}

\begin{figure}
 	\begin{center}
 		\includegraphics[width=.8\textwidth]{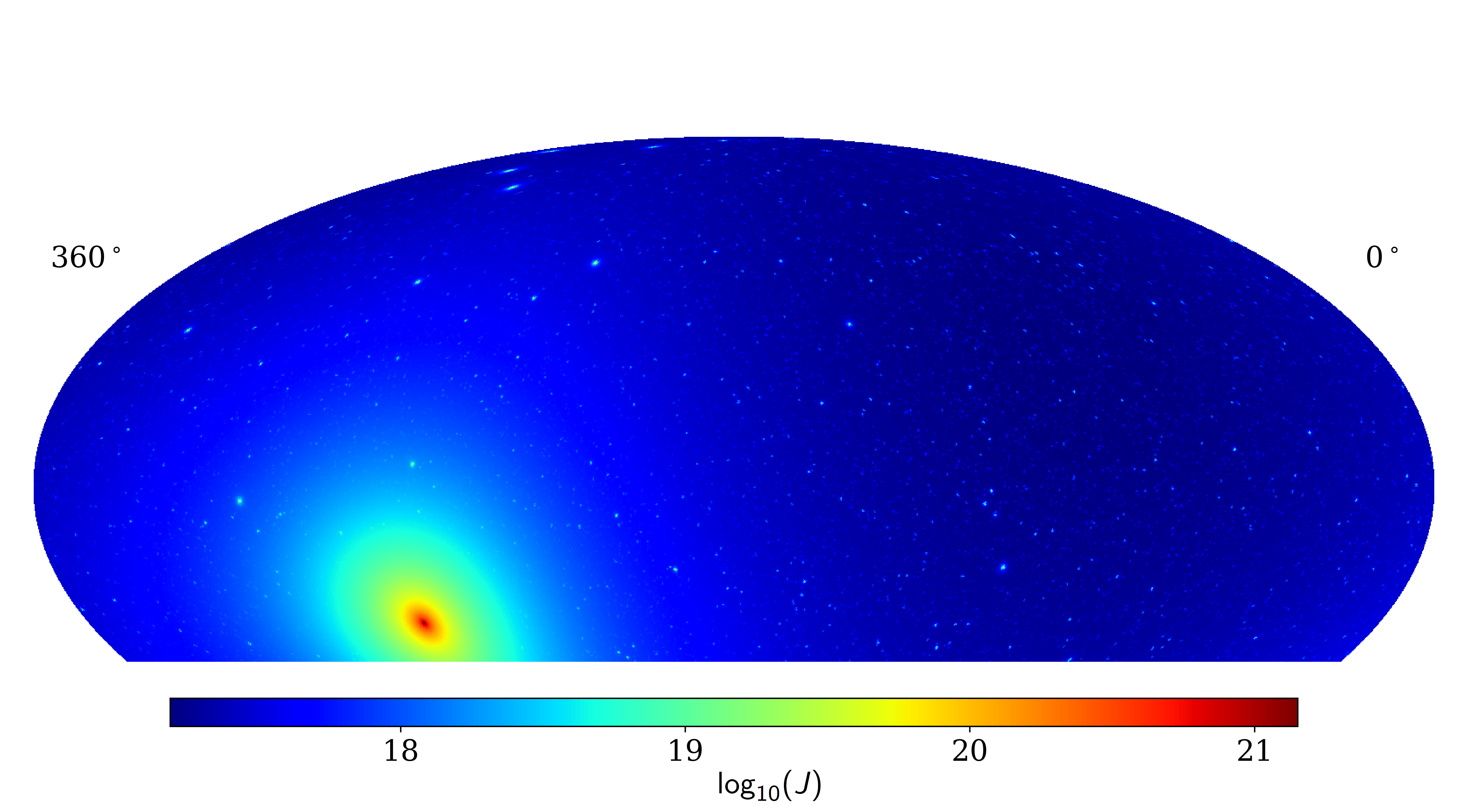}
 		\caption{Sample map showing one simulation of the smooth halo and substructure of the Milky Way halo.  This map is in celestial coordinates and is truncated so as to correspond the HAWC field-of-view. The color scale shows the logarithm of the J-factor (Eq.~\ref{j}) integrated over a single pixel width of solid angle (0.082 $\deg^{2}$).  The high J-factor at the Galactic Center is clearly visible with substructure distributed throughout the sky.}
 		\label{j-map}
 	\end{center}
 \end{figure}

 A sample simulation is shown in Fig \ref{j-map}. This map was generated assuming an Einasto profile for the main halo and sub-halos and is one of the 100 trials used.  Under this profile assumption, we expect to observe roughly 50\% more high J-factor ($>10^{19}~\mathrm{GeV^2~cm^{-5}~sr}$) dwarfs than have been observed.

\subsection{Detection Thresholds}
\label{thresholds}
For each \textsc{clumpy} trial, we compute the HAWC sensitivity to gamma-ray signals from the generated J-factor map of the sky.  
As the HAWC sensitivity is dependent on declination, we bin the sky sensitivity into declination bands of five degrees.   For each dark dwarf we generate the corresponding TS (Eq.~\ref{ts}) profile for a given mass and channel as a function of assumed $\langle \sigma v \rangle$, using the observed background from a randomly chosen pixel within the corresponding declination band. 

 We then calculate  $\langle \sigma v \rangle$ such that at least one of these simulated sub-halos' TS value changes by 25 from the background-only case.  This definition matches that used in the 2HWC catalog (i.e. a source is detected if TS~$\geq$~25 \cite{cat}.)
See Appendix A of Ref. \cite{dwarf} for additional details on evaluating TS values with HAWC.

\begin{figure}[H]
	\begin{center}
		\includegraphics[width=.9\textwidth]{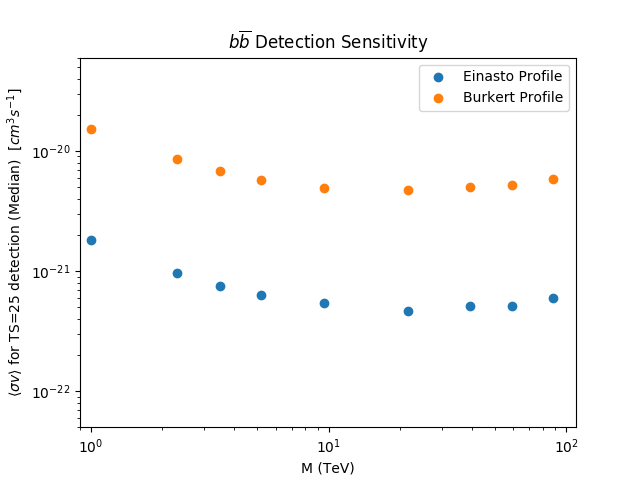}
		\includegraphics[width=.9\textwidth]{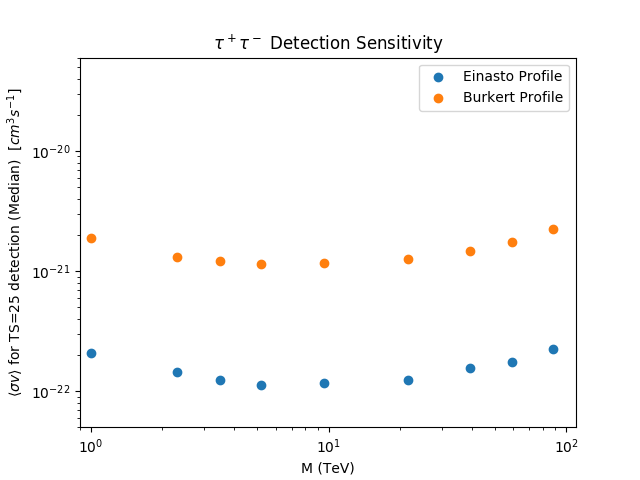}
		\caption{Estimated HAWC detection threshold $\langle \sigma v \rangle$ values for a sample of dark matter simulated substructure.  Substructure models for a range of dark matter particle masses, annihilation channels and halo mass models. are shown. The plotted curves show the $ \langle \sigma v \rangle $ such that  50\% of possible realizations yielded a TS of 25 or greater for at least one dark dwarf.}
		\label{sens} 
	\end{center}
\end{figure}

We repeat the process described above for each of the one-hundred \textsc{clumpy} trials and obtain an ensemble of $\langle \sigma v \rangle$ sensitivity values. In Fig.~\ref{sens}, we report the median of these values as a function of dark matter mass and annihilation channel. The data in Fig.~\ref{sens} should then be interpreted as the $\langle \sigma v \rangle$ value such that HAWC would have a 50\% chance of detecting dark dwarf emission at TS~=~25 within the ensemble of possible realizations of dark matter substructure.

It should be noted that these results are not intended to be as constraining as the 95\% upper limits shown for a hypothetical new dwarf in Fig.~\ref{scaled}.  Resolving a source at TS~=~25 requires
a much larger $\langle \sigma v \rangle$ than can be excluded at TS~=~2.7 (the corresponding number for 95\% one-sided limits). 

In the background-only case, ignoring effects from fluctuations, TS is approximately proportional to $\langle \sigma v \rangle^2$, which can be seen by performing a Taylor expansion on the TS as a function of scale factor (Eq.~A4 of Ref.~\cite{dwarf}).  Therefore, one would expect the sensitivity for this analysis to be a factor of approximately $\sqrt{25}/\sqrt{2.7}\approx3.08$ less constraining than the 95\% CL threshold if considered with the same declination and J-factor.  

To verify this expectation we examine the results for one of our one-hundred \textsc{clumpy} trials, assuming a dark matter mass of 10 TeV.  In this trial  the  $\langle \sigma v \rangle$ threshold was \\$6.06\times 10^{-22}~\mathrm{cm^{-3}s^{-1}}$, and resulted in a TS~=~25 detection of a simulated dwarf with \\$J=10^{19.23}~ \mathrm{GeV^{2} ~cm^{-5} ~sr}$ at a declination of five degrees.  The corresponding characteristic upper limit from Fig \ref{limits}, scaled by this J-factor gives 95\% CL limit on  $\langle \sigma v \rangle$ of  $1.94 \times 10^{-22}~\mathrm{cm^{-3}~s^{-1}}$.  The TS~=~25 case is a factor of 3.1 times less constraining than the 95\% CL case, matching the estimated change.

\subsection{J-factor Sensitivity}
Another possible interpretation of the results in Sec \ref{thresholds} is to determine the minimum J-factor required for a dark dwarf to be detected  (assuming some $\langle \sigma v \rangle$ value and dark matter model).   Rather than associating each likelihood profile with a simulated J-factor, one could instead fix the value of  $\langle \sigma v \rangle$ and compute J such that the TS reaches 25.  We show a sample calculation here for the $ \tau^+ \tau^- $ channel assuming a dark matter mass of 10 TeV (the best-constrained channel according to Fig.~\ref{limits}), and a velocity-weighted cross-section of $10^{-24}~\mathrm{cm^3~s^{-1}}$ (roughly the highest value consistent with the upper limits set in Ref.~\cite{dwarf}).    
  
Using TS profiles generated at a declination of 20 (our most sensitive declination bin), we find the average J-factor required for a TS of 25 to be $5.79\times 10^{20}~\mathrm{GeV^2~cm^{-5}~sr}$.  For comparison the Segue 1 dwarf spheroidal galaxy was found to have a J-factor of $1.1 \times 10^{19} ~\mathrm{GeV^2~cm^{-5} ~sr}$ \cite{magic_segue}.  Given that Segue 1 is at a distance of  23 kpc, and that the J-factor is approximately proportional to the square of distance, a dwarf of comparable mass and scale would need to at most $\sim 3$~kpc away to guarantee a detection by HAWC.
  
\section{Conclusion}
We were able to use the excellent survey capabilities of HAWC to perform an unbiased search for dark matter annihilation originating from substructure.  We observed no significant excesses that could not be equally well explained emission from luminous matter. Though this does not rule out the possibility of the signals originating from dark matter, the HAWC data does not significantly favor the dark matter hypothesis over other astrophysical hypotheses.

We then calculated characteristic limits on flux from dark matter as a function of declination - the sensitivity of HAWC to set limits from discovered dwarf galaxies. Since we did not assume dark matter density profiles for the hypothetical dwarfs, these characteristic limits were done for $J \langle \sigma v \rangle$ rather than $\langle \sigma v \rangle$.  These characteristic limits match what is expected from the HAWC sensitivity, with the most constraining limits coming from the points which transit directly overhead of HAWC.  As new dwarf galaxies are discovered in our survey region, we will be able to estimate expected constraints on $\langle \sigma v \rangle$ by scaling our current limits by the measured J-factors. (Dark matter limits from any individual dwarf are dependent on statistical background fluctuations at the dwarf location.)

With these calculations, we computed the HAWC sensitivity to gamma ray emission from dark dwarfs.  Using simulated realizations of dark matter sub-halos, we found the $\langle \sigma v \rangle$ value which would give an expectation of at least one dark dwarf at TS~=~25 50\% of the time.  Although the ensuing detection thresholds appear less sensitive than the 95\% CL limits shown in previous HAWC analyses, they are consistent with what is expected for the more stringent TS~=~25 criterion used.  A future analysis could improve this sensitivity by taking a combined likelihood approach, finding a total change in TS of 25 above background, summed over all sub-halo contributions.   The combined additional significance above background could reach TS~=~25 at a much lower $\langle \sigma v \rangle$.  In addition, HAWC is continuously taking additional data and improving its reconstruction algorithms. With these additional tools, it will be possible to gain even more sensitivity to dark matter substructure in the gamma-ray regime.

\acknowledgments

We acknowledge the support from: the US National Science Foundation (NSF); the US Department of Energy Office of High-Energy Physics; 
the Laboratory Directed Research and Development (LDRD) program of Los Alamos National Laboratory; 
Consejo Nacional de Ciencia y Tecnolog\'{\i}a (CONACyT), M{\'e}xico (grants 271051, 232656, 260378, 179588, 239762, 254964, 271737, 258865, 243290, 132197, 281653)(C{\'a}tedras 873, 1563, 341), Laboratorio Nacional HAWC de rayos gamma; 
L'OREAL Fellowship for Women in Science 2014; 
Red HAWC, M{\'e}xico; 
DGAPA-UNAM (grants IG100317, IN111315, IN111716-3, IA102715, IN109916, IA102917, IN112218); 
VIEP-BUAP; 
PIFI 2012, 2013, PROFOCIE 2014, 2015; 
the University of Wisconsin Alumni Research Foundation; 
the Institute of Geophysics, Planetary Physics, and Signatures at Los Alamos National Laboratory; 
Polish Science Centre grant DEC-2014/13/B/ST9/945, DEC-2017/27/B/ST9/02272; 
Coordinaci{\'o}n de la Investigaci{\'o}n Cient\'{\i}fica de la Universidad Michoacana; Royal Society - Newton Advanced Fellowship 180385. Thanks to Scott Delay, Luciano D\'{\i}az and Eduardo Murrieta for technical support.

\bibliography{dm_bibliography}

\end{document}

%% file: authors027.tex
\author[a]{A.U.~Abeysekara}

\author[b]{A.~Albert}

\author[c]{R.~Alfaro}

\author[d]{C.~Alvarez}

\author[d]{R.~Arceo}

\author[e]{J.C.~Arteaga-Velázquez}

\author[c]{D.~Avila Rojas}

\author[f]{H.A.~Ayala Solares}

\author[c]{E.~Belmont-Moreno}

\author[g]{S.Y.~BenZvi}

\author[h]{C.~Brisbois}

\author[d]{K.S.~Caballero-Mora}

\author[i]{A.~Carramiñana}

\author[j]{S.~Casanova}

\author[k]{J.~Cotzomi}

\author[i]{S.~Coutiño de León}

\author[k]{C.~De León}

\author[l]{E.~De la Fuente}

\author[m]{S.~Dichiara}

\author[b]{B.L.~Dingus}

\author[n]{M.A.~DuVernois}

\author[l]{J.C.~Díaz-Vélez}

\author[o]{K.~Engel}

\author[c]{C.~Espinoza}

\author[h]{H.~Fleischhack}

\author[m]{N.~Fraija}

\author[m]{A.~Galván-Gámez}

\author[c]{J.A.~García-González}

\author[m]{M.M.~González}

\author[o]{J.A.~Goodman}

\author[b]{J.P.~Harding}

\author[h]{B.~Hona}

\author[d]{F.~Hueyotl-Zahuantitla}

\author[h]{P.~Hüntemeyer}

\author[m]{A.~Iriarte}

\author[p]{A.~Lara}

\author[m]{W.H.~Lee}

\author[c]{H.~León Vargas}

\author[q]{J.T.~Linnemann}

\author[i]{A.L.~Longinotti}

\author[r]{G.~Luis-Raya}

\author[q]{J.~Lundeen}

\author[f]{K.~Malone}

\author[q]{S.S.~Marinelli}

\author[k]{O.~Martinez}

\author[o]{I.~Martinez-Castellanos}

\author[s]{J.~Martínez-Castro}

\author[t]{J.A.~Matthews}

\author[u]{P.~Miranda-Romagnoli}

\author[k]{E.~Moreno}

\author[f]{M.~Mostafá}

\author[j]{A.~Nayerhoda}

\author[v]{L.~Nellen}

\author[a]{M.~Newbold}

\author[g]{M.U.~Nisa}

\author[u]{R.~Noriega-Papaqui}

\author[r]{E.G.~Pérez-Pérez}

\author[t]{Z.~Ren}

\author[g]{C.D.~Rho}

\author[o]{C.~Rivière}

\author[i]{D.~Rosa-González}

\author[f]{M.~Rosenberg}

\author[k]{H.~Salazar}

\author[j]{F.~Salesa Greus}

\author[c]{A.~Sandoval}

\author[o]{M.~Schneider}

\author[b]{G.~Sinnis}

\author[o]{A.J.~Smith}

\author[a]{R.W.~Springer}

\author[q]{K.~Tollefson}

\author[i]{I.~Torres}

\author[w]{G.~Vianello}

\author[n]{T.~Weisgarber}

\author[n]{J.~Wood}

\author[g]{T.~Yapici}

\author[x]{A.~Zepeda}

\author[b]{H.~Zhou}

\author[e]{J.D.~Álvarez}

\affiliation[a]{Department of Physics and Astronomy, University of Utah, Salt Lake City, UT, USA }

\affiliation[b]{Physics Division, Los Alamos National Laboratory, Los Alamos, NM, USA }

\affiliation[c]{Instituto de F\'{i}sica, Universidad Nacional Autónoma de México, Ciudad de Mexico, Mexico }

\affiliation[d]{Universidad Autónoma de Chiapas, Tuxtla Gutiérrez, Chiapas, México}

\affiliation[e]{Universidad Michoacana de San Nicolás de Hidalgo, Morelia, Mexico }

\affiliation[f]{Department of Physics, Pennsylvania State University, University Park, PA, USA }

\affiliation[g]{Department of Physics \& Astronomy, University of Rochester, Rochester, NY , USA }

\affiliation[h]{Department of Physics, Michigan Technological University, Houghton, MI, USA }

\affiliation[i]{Instituto Nacional de Astrof\'{i}sica, Óptica y Electrónica, Puebla, Mexico }

\affiliation[j]{Institute of Nuclear Physics Polish Academy of Sciences, PL-31342 IFJ-PAN, Krakow, Poland }

\affiliation[k]{Facultad de Ciencias F\'{i}sico Matemáticas, Benemérita Universidad Autónoma de Puebla, Puebla, Mexico }

\affiliation[l]{Departamento de F\'{i}sica, Centro Universitario de Ciencias Exactase Ingenierias, Universidad de Guadalajara, Guadalajara, Mexico }

\affiliation[m]{Instituto de Astronom\'{i}a, Universidad Nacional Autónoma de México, Ciudad de Mexico, Mexico }

\affiliation[n]{Department of Physics, University of Wisconsin-Madison, Madison, WI, USA }

\affiliation[o]{Department of Physics, University of Maryland, College Park, MD, USA }

\affiliation[p]{Instituto de Geof\'{i}sica, Universidad Nacional Autónoma de México, Ciudad de Mexico, Mexico }

\affiliation[q]{Department of Physics and Astronomy, Michigan State University, East Lansing, MI, USA }

\affiliation[r]{Universidad Politecnica de Pachuca, Pachuca, Hgo, Mexico }

\affiliation[s]{Centro de Investigaci\'on en Computaci\'on, Instituto Polit\'ecnico Nacional, M\'exico City, M\'exico.}

\affiliation[t]{Dept of Physics and Astronomy, University of New Mexico, Albuquerque, NM, USA }

\affiliation[u]{Universidad Autónoma del Estado de Hidalgo, Pachuca, Mexico }

\affiliation[v]{Instituto de Ciencias Nucleares, Universidad Nacional Autónoma de Mexico, Ciudad de Mexico, Mexico }

\affiliation[w]{Department of Physics, Stanford University: Stanford, CA 94305–4060, USA}

\affiliation[x]{Physics Department, Centro de Investigacion y de Estudios Avanzados del IPN, Mexico City, DF, Mexico }